\documentclass[12pt]{article}
\usepackage{amsfonts,amssymb}

\def\p{ \partial }

\def\bq{ \begin{equation} }
\def\eq{ \end{equation} }
\def\ben{ \begin{eqnarray} }
\def\en{ \end{eqnarray} }

\def\frac#1#2{{#1\over #2}}

\def\on#1#2{\mathop{\vbox{\ialign{##\crcr\noalign{\kern2pt}
$\scriptstyle{#2}$\crcr\noalign{\kern2pt\nointerlineskip}
\kern-2pt$\hfil\displaystyle{#1}\hfil$\crcr}}}\limits}

\begin{document}

\baselineskip=15pt
\vspace{1cm} \centerline{{\LARGE \textbf {Classification of
integrable
 }}}
\vskip0.2cm \hfill
 \centerline{{\LARGE \textbf {Vlasov-type equations   ~~~~
 }}}

\vskip1cm \hfill
\begin{minipage}{13.5cm}
\baselineskip=15pt {\bf A.V. Odesskii ${}^{1,}{}^{2}$, M.V. Pavlov
${}^{3}$, V.V. Sokolov ${}^{2}$}
\\ [2ex] {\footnotesize
${}^{1}$  L.D. Landau Institute for Theoretical Physics (Russia)
\\
${}^{2}$  University of Manchester (UK)
\\
${}^3$ P.N. Lebedev Physical Institute (Russia)
\\}
\vskip1cm{\bf Abstract}
Classification of integrable Vlasov-type equations is reduced
to a functional equation for a generating function. A general
solution of this functional equation is found in terms of hypergeometric
functions.

\end{minipage}

\vglue1cm \textbf{Address}: L.D. Landau Institute for Theoretical
Physics of Russian Academy of Sciences, Kosygina 2, 119334,
Moscow, Russia

\textbf{E-mail}:  odesskii@itp.ac.ru, sokolov@itp.ac.ru
\newpage

\section{Introduction}

Consider the linear equation
\begin{equation} \label{maineq}\Lambda _{t}=\{\Lambda ,\psi
\},\end{equation} where
$\{\Lambda ,\psi \}=\psi _{p}\Lambda _{x}-\psi _{x} \Lambda _{p}.$
Here the unknown function $\Lambda(x,t,p)$ depends on  the
 ``spectral'' parameter $p$ and a function $\psi=\psi (U(x,t),p)$.
 Following \cite{zakh}, we call (\ref{maineq}) {\it the Vlasov-type equation} generated by $\psi(U,p).$
 Equation (\ref{maineq}) is also known as the dispersionless Lax equation.

It is easy to check that a partial hodograph transformation
$\Lambda(x,t,p)\rightarrow p(x,t,\Lambda)$ reduces (\ref{maineq})
to the following conservative form
\begin{equation}  \label{psi}
p_t=\psi(U,p)_x.
\end{equation}
Here $\Lambda$ plays a role of parameter.

For some functions $\psi$ Vlasov-type equations are closely
related to integrable hydrodynamic chains \cite{pav1,pav2}.
A hydrodynamic chain associated with Vlasov-type equation can be
derived by expanding $\Lambda(x,t,p)$ at a singular point
of the function $\psi$. In such a case, formula
(\ref{psi}) yields conservation laws for the hydrodynamic chain.

{\bf Example 1}
 (\textit{The Benney chain}) \cite{benney, gibbons, kupman, zakh}.
The Vlasov equation (or collisionless Boltzmann equation) has the form
\begin{equation}
\Lambda _{t}+\Lambda _{p}U_{x}-p\Lambda _{x}=0,  \label{10}
\end{equation}
where $\psi=\frac{p^2}{2}+U.$ Substituting the expansion
\begin{equation}
\Lambda =p+\frac{A^{0}}{p}+\frac{A^{1}}{p^{2}}+\frac{A^{2}}{p^{3}}+\frac{
A^{3}}{p^{4}}+...  \label{11}
\end{equation}
into (\ref{10}), one derives the famous Benney hydrodynamic chain
$$
A_{t}^{k}=A_{x}^{k+1}+kA^{k-1}A_{x}^{0},\qquad k=0,1,2,...,
$$
where $A^0=U.$  Let
$$
 p= \Lambda-\frac{H_{0}}{\Lambda}-\frac{H_{1}}{\Lambda^{2}}-\frac{H_{2}}{\Lambda^{3}}-\frac{
H_{3}}{\Lambda^{4}}-...
$$ be the inverse series for (\ref{11}). Functions $H_i$ can be easily calculated:
$H_0=A^0, H_1=A^1, H_2=A^2+(A^0)^2, \dots$. The formula
$$
p_t= \Big(\frac{p^2}{2}+U \Big)_x
$$
generates infinitely many conservation laws for the Benney chain:
$$
(H_k)_t=\Big(H_{k+1}-\frac12 \sum_{i=0}^{k-1} H_i H_{k-i-1} \Big)_x.
$$

{\bf Example 2}  (\textit{The Kupershmidt chain}) \cite{kuper, pav3}.
The Kupershmidt hydrodynamic chain
$$
B_{t}^{k}=B_{x}^{k+1}+B^{0}B_{x}^{k}+\beta k B^{k}B_{x}^{0},\qquad k=0,1,2,...
$$
is connected to the Vlasov-type equation (\ref{maineq}), where
$$
 \psi=  \frac{p^{\beta +1}}{\beta +1}+U\, p,
$$
by the expansion
$$
\Lambda =p^{\beta}+B^{0}+\frac{B^{1}}{p^{\beta}}+\frac{B^{2}}{p^{2
\beta}}+\frac{B^{3}}{p^{3 \beta}}+...,
$$
where $B^0=U.$ Conservation laws for this chain can be calculated
in the same way as in Example 1.

The hydrodynamic chains described in Examples 1, 2 admit infinitely
many hydrodynamic reductions \cite{ferhus1, Gibt}.
The corresponding Vlasov-type equations admit the same reductions.

The following ``integrable'' functions $\psi$ were found in \cite{pav1}:

\textbf{Case 1}. $\psi(U,p)=U+W(p)$, where
$$
W''=c_1 W'^2+c_2 W'+c_3
$$
and

 \textbf{Case 2}. $\psi(U,p)=U\,W(p)$, where
$$
W''=\frac{1}{W} \Big( c_1 W'^2+c_2 W'+c_3 \Big).
$$
 Here $c_i$ are arbitrary parameters.  The
Benney chain corresponds to Case 1 with $W(p)=p^2/2.$

In this paper we describe all possible ``integrable'' functions
$\psi(U, p)$ using the method of hydrodynamic reductions.
The existence of hydrodynamic reductions have been proposed as a definition
of integrability for dispersionless multi-dimensional equations
in \cite{ferhus1}. We apply this approach for
Vlasov-type equations.

\section{Hydrodynamic reductions}

Suppose there exists a semi-Hamiltonian \cite{tsar} hydrodynamic-type
system
\begin{equation}\label{gidra}
r_{t}^{i}=v^{i}(\mathbf{r})r_{x}^{i}\qquad i=1,2,...,N,
\end{equation}
and functions $U=u(\mathbf{r})$ and
$\Lambda=\lambda(\mathbf{r},p)$ such that these functions satisfy
(\ref{maineq}) for any solution ${\bf r}(x,t)$ of system
(\ref{gidra}). Then (\ref{gidra}) is called
  {\it a hydrodynamic reduction} for the Vlasov-type equation
(\ref{maineq}). The partial hodograph transformation
$\lambda(\mathbf{r},p)\rightarrow p({\bf r},\lambda)$ leads to the
corresponding hydrodynamic reduction of (\ref{psi}).

Substituting $\Lambda=\lambda(\mathbf{r},p)$ and $U=u(\mathbf{r})$
in (\ref{maineq}) and (\ref{psi}), we obtain the equations
 \begin{equation}
 \lambda _{t}=\psi_p \lambda_x-\psi_x \lambda_p  \label{lamt}
 \end{equation}
 and
 \begin{equation}  \label{psi1}
p_t=\psi(u,p)_x.
\end{equation}
Calculating the derivatives by virtue of  (\ref{gidra}), we obtain from
(\ref{lamt}) that
$$
\sum \Big(\lambda _{p}\psi _{u}\, \partial _{i} u+[v^{i}(\mathbf{r})- \psi _{p}]\partial
 _{i}\lambda\Big)\, r_x^i=0 ,
$$
where we use the notation $\partial _{i}=\partial /\partial
r^{i}$. Since $\bf r$ is {\it arbitrary} solution of system
(\ref{gidra}), we get
\begin{equation}
\lambda _{p}\psi _{u}\, \partial _{i} u=[\psi
_{p}-v^{i}(\mathbf{r})]\,\partial _{i}\lambda , \qquad i=1,2,...,N.
\label{6}
\end{equation}

Let us determine functions $p^{i}(\mathbf{r}),$
$i=1,2,...,N$ as solutions of the equations
\begin{equation}
v^{i}(\mathbf{r})=\psi _{p}|_{p=p^{i}}.  \label{5}
\end{equation}
Then (\ref{6}) implies that the equation $ \lambda _{p}=0 $
has $N$ solutions (pairwise distinct in the generic case), i.e.
\begin{equation}
\lambda _{p}|_{p=p^{i}}=0,\qquad i=1,2,...,N.  \label{1}
\end{equation}
Without loss of generality we can fix the Riemann invariants $r^i$ of the
system (\ref{gidra}) in such a way that
\begin{equation} \label{ri} r^{i}=\lambda |_{p=p^{i}}.\end{equation}
Indeed, if we substitute $p=p^i$ into equation (\ref{lamt}), then (\ref{lamt}), (\ref{1})
imply
$$
( \lambda|_{p=p^{i}})_t=(\psi_p|_{p=p^{i}})   ( \lambda|_{p=p^{i}})_x.
$$
This means (see (\ref{5})) that  $\lambda|_{p=p^{i}}$ satisfies
(\ref{gidra}) and therefore  $\lambda|_{p=p^{i}}=R_i(r^i)$ for
some functions $R_i.$ According to (\ref{1}), the branch points of the
Riemann surface determined by the equation
 $ \Lambda =\lambda (\mathbf{r},p),$ are nothing but the Riemann invariants of system (\ref{gidra}).
 This fact is well-known for hydrodynamic-type systems that produced by the Whitham averaging
 procedure applied to multi-phase solutions of both integrable
 continuous dispersive equations and integrable discrete equations (see references in \cite{pav2}).


Substituting functions $p({\bf r}, \lambda)$,  $ u({\bf r})$ in
(\ref{psi1}), we obtain
\begin{equation} \label{qq0}\partial _{i}p=\frac{\psi _{u}\partial _{i}u}{\psi _{p}|_{p=p^{i}}-\psi
_{p}}.\end{equation} If we fix $\lambda=r^k, k\ne i,$ then
(\ref{ri}) implies $p=p^k$ and we obtain \begin{equation}
\label{qq1}
\partial _{i}p^{k}=\frac{\psi _{u}|_{p=p^{k}}\partial _{i} u}{\psi_{p}|_{p=p^{i}}-\psi
_{p}|_{p=p^{k}}}.\end{equation} Let us introduce the following
notation:
$$
f_i=\frac{\psi_u}{\psi_p\vert_{p=p^{i}}-\psi_p}, \qquad \quad
f_{ik}=\frac{\psi_u\vert_{p=p^{k}}}{\psi_p\vert_{p=p^{i}}-\psi_p\vert_{p=p^{k}}},
\quad i\ne k.
$$
The compatibility conditions
 $\partial _{k}(\partial _{i}p)=\partial
_{i}(\partial _{k}p),\quad i\neq k
$
are equivalent to the equations
\begin{equation} \label{qq2}
\p_{ik}^{2}u=\frac{f_{ik} \,\p_{p^{k}} f_k-f_{ki} \,\p_{p^{i}}
f_i+\p_u (f_k-f_i)+f_i \, \p_p f_k-f_k \,\p_p f_i}{f_i-f_k}
\,\,\partial _{i} u \,\partial _{k} u.\end{equation} Equations
(\ref{qq1}), (\ref{qq2}) form a system of equations named in \cite{pav1} {\it the generalized Gibbons--Tsarev system}
(cf. \cite{Gibt}).

Since $u$ does not depend on $p,$
(\ref{qq2}) implies the following functional equation
\begin{equation}   \label{funeq}
\p_p \Big(
\frac{f_{ik} \,\p_{p^{k}} f_k-f_{ki} \,\p_{p^{i}} f_i+\p_u
(f_k-f_i)+f_i \, \p_p f_k-f_k \,\p_p f_i}{f_i-f_k}
\Big)=0
\end{equation}
for the function $\psi(u,p).$ In the next sections we study this
functional equation and found its general solution.
The general solution is expressed in terms of a pair of arbitrary solutions of the standard
hypergeometric equation
\begin{equation}   \label{hypeq}
u (u-1) \, y(u)''+ [(\alpha+\beta+1)\,u-\gamma]\, y(u)'+\alpha
\beta\, y(u)=0.
\end{equation}
Note that the general solution of the Chazy equation, which appears in the
classification paper \cite{pav4}, can also be parameterized by a pair
of hypergeometric functions. Our solution $\psi(u,p)$ is a
generalization of the solution $h(\xi,u)$ found in \cite{odsok}
(see Example 4).

\section{Particular solutions}

In Sections 4, 5 we solve the functional equation (\ref{funeq}) in terms of quadratures of
hypergeometric functions. In this section we consider some particular cases, where
the result can be written more explicitly.

Computing the numerator of the left hand side of (\ref{funeq}) and
expanding it at $p^i=p, \, p^k=p,$ we obtain that the
vanishing of the coefficients up to 8-th degree in the corresponding
Taylor series is equivalent to the following system of PDEs
for the function $\psi(u,p)$:
\begin{equation}\begin{array}{l}\label{eq1}
3 \psi_{ppp}^3 \psi_u^3-4 \psi_{pp} \psi_{ppp} \psi_{pppp}
\psi_u^3+\psi_{pp}^2 \psi_{ppppp} \psi_u^3-3  \psi_{pp}
\psi_{ppp}^2 \psi_u^2 \psi_{pu}+2 \psi_{pp}^2 \psi_{pppp} \psi_u^2
\psi_{pu}+\\[4mm]
6\, \psi_{pp}^2 \psi_{ppp} \psi_u^2 \psi_{ppu}-5 \psi_{pp}^3
\psi_u^2 \psi_{pppu}-6 \psi_{pp}^4 \psi_{pu} \psi_{uu}+6
\psi_{pp}^4 \psi_{u} \psi_{puu}=0,
\end{array}
\end{equation}

\begin{equation}\begin{array}{l}
3 \psi_{ppp}^2 \psi_u^3 \psi_{ppu}-\psi_{pp} \psi_{pppp} \psi_u^3
\psi_{ppu}-3 \psi_{pp} \psi_{ppp} \psi_u^3 \psi_{pppu}+
\psi_{pp}^2
  \psi_u^3 \psi_{ppppu}-3 \psi_{pp}^2 \psi_{ppp} \psi_u \psi_{pu}
  \psi_{uu}-
\\[4mm]
6\, \psi_{pp}^3 \psi_{pu}^2 \psi_{uu}+3 \psi_{pp}^3 \psi_u
\psi_{ppu} \psi_{uu}+3 \psi_{pp}^2 \psi_{ppp} \psi_{u}^2
\psi_{puu}+6 \psi_{pp}^3 \psi_{u} \psi_{pu} \psi_{puu}-3
\psi_{pp}^3 \psi_{u}^2 \psi_{ppuu}=0, \label{eq2}
\end{array}
\end{equation}
and
\begin{equation}\begin{array}{l}\label{eq3}
-3 \psi_{ppp}^2 \psi_u^2 \psi_{pu} \psi_{uu}+
\psi_{pp}\psi_{pppp}\psi_u^2 \psi_{pu} \psi_{uu} -6 \psi_{pp}
\psi_{ppp}  \psi_u  \psi_{pu}^2 \psi_{uu} -6 \psi_{pp}^2
\psi_{pu}^3 \psi_{uu}+
\\[4mm]
3\, \psi_{pp} \psi_{ppp} \psi_u^2 \psi_{ppu} \psi_{uu}+6
\psi_{pp}^2 \psi_u  \psi_{pu} \psi_{ppu} \psi_{uu} - \psi_{pp}^2
\psi_u^2 \psi_{pppu} \psi_{uu}+3 \psi_{ppp}^2 \psi_{u}^3
\psi_{puu}-
\\[4mm]
\psi_{pp} \psi_{pppp} \psi_u^3 \psi_{puu}+6\, \psi_{pp} \psi_{ppp}
\psi_u^2 \psi_{pu} \psi_{puu} +6\, \psi_{pp}^2 \psi_u \psi_{pu}^2
\psi_{puu} -3\, \psi_{pp}^2 \psi_{u}^2 \psi_{ppu} \psi_{puu}-
\\[4mm]
3\,\psi_{pp} \psi_{ppp} \psi_u^3 \psi_{ppuu}-3\, \psi_{pp}^2
\psi_u^2 \psi_{pu} \psi_{ppuu} + \psi_{pp}^2 \psi_u^3
\psi_{pppuu}=0.
\end{array}
\end{equation}
This system and the functional equation (\ref{funeq}) are
invariant under any transformation of the form $u\rightarrow f(u).$ They are also invariant with
respect to the following symmetry group:
\begin{equation}\label{tran}
\psi\rightarrow c_2 \psi+c_1 p+c_0, \qquad p\rightarrow k_2 \psi+k_1
p+k_0.
\end{equation}
These symmetries are associated with linear transformations of
independent variables in (\ref{maineq}), (\ref{psi}). Notice that any
function $\psi=W(u)+p\, V(u)$ satisfies equation (\ref{funeq}) and system
(\ref{eq1})-(\ref{eq3}).

The integrable cases $\psi(u,p)=u+W(p)$ and $\psi(u,p)=u\, W(p)$
described in Introduction can be found directly from (\ref{eq1}).
We denote these cases as Case 1 and Case 2, respectively. We mention two
more particular integrable cases:

\textbf{Case 3}. $\psi(u,p)=W(u-p)$, where
$$
W''=c_1 W'^3+c_2 W'^2+c_3 W' ;
$$

\textbf{Case 4}. $\psi(u,p)=p\, u+W(p)$, where
$$
W''=\frac{1}{p} \Big( c_1 W'^2+c_2 W'+c_3 \Big).
$$
Here $c_i$ are arbitrary constants. The Kuperhsmidt chain belongs to Case
4. It is easy to see that Case 3 is connected to Case 1 by the
transformation $\psi \leftrightarrow p$. In all these four
particular cases  system (\ref{eq1})-(\ref{eq3}) is equivalent to
some ordinary differential equation of the fifth order  for the function $W.$

System (\ref{eq1})-(\ref{eq3}) admits the substitution
\begin{equation} \label{sub}
\psi_p=F(p, \psi),
\end{equation}
corresponding to the factorization with respect to the symmetry group
$u\rightarrow f(u).$ As the result, one gets an overdetermined system
of three PDEs for the function $F(x,y).$  We do not present this system here because
of its complexity.

Case {\bf 1} corresponds to the case $\bar{\bf 1}$: $F(x,y)=V(x)$,
where $V=W'$. Case {\bf 3} corresponds to $\bar{\bf 3}$:
$F(x,y)=V(y)$. Case {\bf 2} corresponds to $\bar{\bf 2}$:
$F(x,y)=y V(x)$, where $V=W'/W.$ Case {\bf 4} corresponds to
$\bar{\bf 4}$: $F(x,y)=y/x+ V(x)$, where $V=W'-W/x.$

Another particular solution is determined by the function
\begin{equation}\label{exam}
F(x,y)=\frac{2x+\alpha }{x^{2}+\alpha x+\beta }\cdot\frac{y^{2}+\gamma
y +\delta }{2 y +\gamma }.
\end{equation}
The corresponding equation (\ref{sub}) leads to equation (\ref{psi1}) of the form
$$
p_{t}=\partial _{x}\sqrt{u\,(p^{2}+t_1 )-t_2 },
$$
where $t_i$ are arbitrary constants. In this example the function
$F$ has the form
\begin{equation}   \label{VW}
F(x,y)=\frac {V(y)}{W(x)}.
\end{equation}
An investigation of this ansatz leads to the following result. The
case, where $V$ or $W$ is a linear function, was considered above.
Namely, the cases $W=1$, $V=1$ and $V=y$ coincide with Cases
3, 1 and 2, respectively. The case $W=x$ transforms to Case 2 by the
substitution $\psi \leftrightarrow p$. If $V''\ne 0$ and $W''\ne 0,$
then the following three classes of solutions (\ref{VW}) exist:

\textbf{Case 5}. ~~~~~~~~~~
$VV^{\prime \prime } =2V^{^{\prime ^{2}}}+c_{1}V^{\prime }+c_{2},
\qquad WW^{\prime \prime }2W^{^{\prime ^{2}}}+ c_{1}W^{\prime}+c_{2},
$

\textbf{Case 6}. ~~~~~~~~~~
$
VV^{\prime \prime } =V^{^{\prime ^{2}}}+c_{1}V^{\prime }+c_{2},
\qquad WW^{\prime \prime }=W^{^{\prime ^{2}}}+c_{1}W^{\prime
}+c_{2},
$

\textbf{Case 7}. ~~~~~~~~~~
$
VV^{\prime \prime } =-V^{^{\prime ^{2}}}+c_{1}V^{\prime }+c_{2},
\qquad WW^{\prime \prime }=-W^{^{\prime ^{2}}}+c_{1}W^{\prime
}+c_{2}.
$

\noindent For all these cases the generic solution depends on 6 arbitrary parameters.
Function (\ref{exam}) belongs to Case 5 with $c_1=-3,$ $ c_2=1.$

\section{General solution}

Let us expand the left hand side of (\ref{funeq}) into the Taylor series
at $p^k=p$. Denote by $S(p^i,p)$ the first nontrivial coefficient of this
expansion. For fixed $p$ consider $S=0,$ $ S_p=0$ as a system of linear algebraic equations
with respect to derivatives $\psi_u,
\psi_{u p^i}$. Its determinant does not vanish if $
\psi_u \psi_{pp}\ne 0$. Solving this system, we obtain
\begin{equation}\label{eqq1}
 \psi_u=\frac{Q(\psi_p)}{\psi_{pp}},
 \end{equation}
where $Q$ is a polynomial with respect to $\psi_p$ of degree not greater than 4
with coefficients depending on $u$ only.
Taking into account (\ref{eqq1}), it is easy to extract from the equation $S_{p^i}=0$ that
\begin{equation}\label{eqq2}
\frac{\psi_{ppp}}{\psi_{pp}^2}=\frac{R(\psi_p)}{Q(\psi_p)},
 \end{equation}
where $R$ is a polynomial of degree not greater than three.
 The compatibility condition of equations
(\ref{eqq1}) and (\ref{eqq2}) has the form
\begin{equation}\label{eqq3}
 Q^2 Q_{xxx}-R Q Q_{xx}+R Q_x^2-(R^2+R_x Q) Q_x+R Q_u-R_{xx}
 Q^2+(2 R R_x-R_u) Q=0,
 \end{equation}
where $x=\psi_p.$

Assume that the polynomial $$Q(x)=a (x-b_1)(x-b_2)(x-b_3)(x-b_4)$$
has distinct roots and rewrite (\ref{eqq2}) as
\begin{equation}\label{drob1}
\frac{\psi_{ppp}}{\psi_{pp}^2}=\frac{k_1}{\psi_p-b_1}+...+\frac{k_4}{\psi_p-b_4}.
\end{equation}
One can verify that the system consisting of equations
(\ref{eq1})-(\ref{eq3}) and (\ref{eqq3}) is equivalent to the
following :

\noindent  1. The functions $k_i(u)$ are arbitrary constants such
that $k_1+...+k_4=3$.

\noindent  2. The functions $a(u),$ $ b_i(u)$ satisfy the following
system of ODEs:
\begin{equation}\label{sys1}
b_i'=a (1-k_i) \prod_{j\ne i} (b_i-b_j), \qquad i=1,...,4.
\end{equation}
The function $a(u)$ can be chosen arbitrarily due to the
admissible transformations  $u\rightarrow s(u).$ Consider
the double ratio
$$
\rho=\frac{(b_1-b_2)(b_3-b_4)} {(b_1-b_3)(b_2-b_4)}.
$$
Differentiating $\rho$ by virtue of (\ref{sys1}), it is easy to check that $\rho'\ne 0$.
Let us change $u$ in such a way that
$\rho=u$. This means that we choose
$$
a=\frac{1}{(b_2-b_3)(b_1-b_4)}+\frac{1}{(b_1-b_2)(b_3-b_4)}.
$$
One can verify that the formulae
$$
b_1=\frac{z_2+u y_2}{z_1+u y_1}, \qquad  b_2=\frac{y_2}{y_1}, \qquad  b_3=\frac{z_2+y_2}{z_1+y_1},
\qquad  b_4=\frac{z_2}{z_1},
$$
where $(y_1,z_1),$     $(y_2,z_2)$ are two arbitrary solutions of
the linear system
\begin{equation}\label{sys2}
 y'=\frac{k_1+k_2+k_3-2 }{u-1}\, y+\frac{k_1+k_2+k_3-2 }{u (u-1)} \,z, \qquad
 z'=\frac{1-k_1}{u-1} \,y    +\frac{1-k_1-k_2+k_2 u}{u (u-1)} \,z,
\end{equation}
define a general solution of (\ref{sys1}).  Notice
that if $k_1+k_2+k_3\ne 2,$ then
$$
z=-u y+\frac{u (u-1)}{k_1+k_2+k_3-2} \, y'
$$
and system (\ref{sys2}) is equivalent to the hypergeometric equation
(\ref{hypeq}), where $k_1=1+\alpha-\gamma$, $k_2=1-\alpha, \,$ $
k_3=\gamma-\beta.$

System (\ref{eqq1}), (\ref{eqq2}) can be reduced to quadratures by
the following way. Let us determine a function $\phi(u,p)$ as the
solution of the over-determined system:
\begin{equation} \label{phi}
\phi_u=-\frac{\phi (\phi-1)\,y_1'}{\beta (y_1 \phi+z_1)}, \qquad
\phi_p=\frac{\phi^{k_1} (\phi-u)^{k_2} (\phi - 1)^{k_3}}{y_1
\phi+z_1}.
\end{equation}
It is easy to check that this system is consistent. Then the
solution of the following system in involution
\begin{equation} \label{ans}
\psi_u=\frac{y_2 y_1'-y_1 y_2'}{\beta (y_1 \phi+z_1)} \,\phi^{1-k_1}
(\phi-u)^{1-k_2} (\phi-1)^{1-k_3}, \qquad \psi_p=\frac{y_2
\phi+z_2}{y_1 \phi+z_1},
\end{equation}
is a general solution of (\ref{funeq}). This fact can be verified by a
direct calculation. It turns out that the
expression under differentiating in the left hand side of (\ref{funeq})  is equal to
$$
\frac{(1-u-\beta) \alpha^2+(1-u-\alpha) \beta^2+u \,(4 \alpha \beta
-\alpha-\beta)}{u(u-1) (\alpha-\beta)^2},
$$
where $\alpha=\phi(u,p^i), \beta= \phi(u,p^k)$.

{\bf Remark.} The standard Wronskian formula for second order linear ODE
implies that the expression $y_2
y_1'-y_1 y_2'$ from (\ref{ans}) equals $C u^{k_1+k_2-2}
(u-1)^{k_2+k_3-2}$ for some constant $C$.

\section{Degenerations}

In Section 4 we have considered the general case. This means that the
polynomial $Q$ has degree 4 and all its roots $b_i$ are distinct
for the generic value $u.$  In this
section we consider degenerations. It is easy to see that the degree of the polynomial $Q$
can be fixed by 4 with the help of transformations (\ref{tran}). It turns out that in all
cases the result can be parameterized by a pair of solutions of
some degenerations of the hypergeometric equation.

{\bf Degeneration 1.} Suppose $Q=a (x-b_1)^2 (x-b_2) (x-b_3);$ then
$$
\frac{R}{Q}=\frac{k_1}{x-b_1}+\frac{f_1}{(x-b_1)^2}+\frac{k_2}{x-b_2}+\frac{k_3}{x-b_3},
$$
where $k_1,k_2,k_3$ are constants such that $k_1+k_2+k_3=3$, and
\begin{equation}\label{bb1}
b_1'=-a (b_1-b_2)(b_1-b_3) f_1, \qquad b_2'=a (b_2-b_1)^2 (b_2-b_3)
(1-k_2),$$$$b_3'=a (b_3-b_1)^2 (b_3-b_2) (1-k_3), \qquad  f_1'=a f_1
\Big((b_1-b_2)(b_1-b_3)(2-k_1)+(b_2+b_3-2 b_1) f_1 \Big).
\end{equation}
A general solution of  system (\ref{bb1}) can be parameterized in the
following way:
$$
b_2=\frac{y_1}{y_2}, \qquad b_3=\frac{z_1}{z_2}, \qquad b_1=\frac{y_1+u z_1}{y_2+u z_2},
$$
where $(y_1,z_1),$     $(y_2,z_2)$ are two arbitrary solutions of
the linear system
\begin{equation}\label{sys4}
 y'=\Big(-\frac{1}{2}+\frac{3 k_2+k_1-4 }{2 (k_2-1)\, u} \Big)\, y+  \,z, \qquad
 z'=\frac{k_1+k_2-2}{(1-k_2) \,u^2} \,y    +\Big(-\frac{1}{2}-\frac{3 k_2+k_1-4 }{2 (k_2-1)\, u} \Big)
 \,z.
\end{equation}
The functions $a, f_1 $ are determined by system (\ref{bb1}). Note that system (\ref{sys4})
is equivalent to the Bessel equation
$$
y''+y'+\Big(\frac{1}{4}+\frac{(k_2-k_1)(k_1+k_2-2)}{4 (k_2-1)^2\,u^2} \Big) \,y=0.
$$

{\bf Degeneration 2.} Suppose $Q=a (x-b_1)^2 (x-b_2)^2;$   then
$$
\frac{R}{Q}=\frac{k_1}{x-b_1}+\frac{f_1}{(x-b_1)^2}+\frac{k_2}{x-b_2}+\frac{f_2}{(x-b_2)^2},
$$
where $k_1,k_2$ are constants such that $k_1+k_2=3,$ and
$$
b_1'=-a (b_1-b_2)^2 f_1, \qquad b_2'=-a (b_1-b_2)^2 f_2,$$$$ f_1'=a
(b_2-b_1) f_1 \Big((b_2-b_1)(k_2-1)+2 f_1 \Big), \qquad f_2'=a
(b_1-b_2) f_2 \Big((b_1-b_2)(k_1-1)+2 f_2 \Big).
$$
The general solution is given by
$$
b_1=\frac{y_1}{y_2}, \qquad b_2=\frac{z_1}{z_2},
$$
where $(y_1,z_1),$     $(y_2,z_2)$ are two arbitrary solutions of
the linear system
\begin{equation}\label{sys3}
 y'= \frac{3 k_1-5 }{2\, u}  \, y+ u \,z, \qquad
 z'=-\frac{1}{3} \,y    + \frac{3(1-k_1)}{2  \, u} \,z.
\end{equation}
Notice that the function $y(u)$ satisfies the following
second order equation:
$$
 y''+\Big(\frac{u}{3}-\frac{3(3 k_1-5)(3 k_1-7)}{u^2}  \Big)\,y=0.
$$

{\bf Degeneration 3.} Suppose $Q=a (x-b_1)^3 (x-b_2);$  then
$$
\frac{R}{Q}=\frac{k_1}{x-b_1}+\frac{f_1}{(x-b_1)^2}+\frac{f_2}{(x-b_1)^3}+\frac{k_2}{x-b_2},
$$
where $k_1,k_2$ are constants, such that $k_1+k_2=3,$ and
$$
b_1'=a (b_2-b_1) f_2, \qquad b_2'=a (b_2-b_1)^3 (1-k_2),$$$$ f_1'=a
  \Big((b_2-b_1)(f_1^2-k_2 f_2 )-2 f_1 f_2 \Big), \qquad f_2'=2 a
  f_2 \Big((b_2-b_1)f_1 - f_2 \Big).
$$
The general solution can be written in the following form
$$b_1=\frac{y_1}{y_2},\qquad  b_2=\frac{y_1'+u y_1}{y_2'+u y_2},$$
where $y_1, y_2$ are arbitrary solutions of the Weber equation
$y''=\Big(u^2+(1-2 k_2) \Big) y.$ The functions $a, f_1, f_2$ are
completely determined by the above system.

{\bf Degeneration 4.} Suppose $Q=a (x-b)^4;$ then
$$
\frac{R}{Q}=\frac{k_1}{x-b}+\frac{f_1}{(x-b)^2}+\frac{f_2}{(x-b)^3}+\frac{f_3}{(x-b)^4},
$$
where $k_1=3,$ and
$$
b'=-a f_3, \quad f_3'=-3 a f_2 f_3, \quad  f_2'=-2 a (f_2^2+f_1
f_3), \quad f_1'=a (f_3-2 f_1 f_2).
$$
Eliminating $f_3, f_2, f_1$ from this system and choosing
$a=-1/b'^2,$ one obtains the equation
$$
\frac{b'''}{b'}-\frac{3 }{2 }\frac{ b''^2}{ b'^2}=-2 u.
$$
Its general solution can be written in the form $b=y_1/y_2,$ where
$y_1, y_2$ are arbitrary solutions of the Airy equation $y''=u y.$

Deeper degenerations can be obtained by the restriction
that the polynomial $Q$ possesses one or several constant roots. In
this case it is convenient to make one of these roots the infinity
using transformation (\ref{tran}). Consider, for
instance, Degeneration 4 under assumption $b=const$. Choosing the
normalization $a=1,$ one obtains $f_3=0$ and
$$
f_2'=-2  f_2^2, \qquad f_1'=  -2 f_1 f_2.
$$
The simplest solution $f_2=f_1=0$ of this system  corresponds (for
$b=\infty$) to the Benney pseudo-potential from Example 1. The
solution $f_2=0, f_1\ne 0$ implies  $\psi(u,p)=u p+p \log{(p)},$
which coincides with Case 4 from  Section 3 for $c_1=c_2=0,
c_3=1.$ Finally, if $f_2\ne 0$, one can obtain (up to the
equivalence)
$$
\psi_u \psi_{pp}=1, \qquad \psi_{pp}=\lambda \sqrt{u} \exp{\Big(-\frac{1}{4 u} \psi_p^2\Big)}.
$$
In this case the  solution cannot be expressed in terms of elementary
functions.

For general system (\ref{sys1}) the fact that some roots $b_i$
are constant, is equivalent to the equality $k_i=1$ for
corresponding values $k_i.$ Solutions of system (\ref{sys1}) for
such degenerations can be extracted from (\ref{phi}), (\ref{ans}). We omit the explicit
formulae for such cases and the analysis of the case of constant $b_i$ for
Degenerations 1-3.

Let us describe particular solutions from Section 3 in the context
of Sections 4, 5. It turns out that Case 5 with $c_1^2\ne 8 c_2$ is equivalent to
(\ref{sys1}), where $b_1=-b_2,$ $b_3=-b_4$, $k_2=k_1$ è $k_4=k_3$.
Namely,
$$k_1=\frac{3 c_3-c_1}{4 c_3}, \qquad k_3=\frac{3 c_3+c_1}{4
c_3}.$$ for $c_2=c_1^2/8-c_3^2/8$.

Case 6 with $c_1^2\ne 4 c_2$ is equivalent to (\ref{sys1}), where
two roots $b_i$ are constant. The right hand side of (\ref{eqq2})
reduces to the form
$$
\frac{1}{x} +\frac{k_1}{x-b_1}+ \frac{k_2}{x-b_2},
$$
where $x=\psi_p,$
$$k_1=\frac{c_3-c_1 }{2 c_3}, \qquad
k_2=\frac{c_3+c_1}{2 c_3}.$$ for $c_2=c_1^2/4-c_3^2/4.$

In Case 7 with $c_1^2\ne -4 c_2$ we have $Q(x)=a x (x-b_1) (x-b_2)$ and
\begin{equation}\label{2pol}
\frac{R}{Q}= \frac{k_1}{x-b_1}+ \frac{k_2}{x-b_2},
\end{equation}
where
$$k_1=\frac{3 c_3-c_1 }{2 c_3}, \qquad
k_2=\frac{3 c_3+c_1}{2 c_3}$$ for $c_2=-c_1^2/4+c_3^2/4.$  Notice that in this case $k_1+k_2=3.$
It is easy to verify that
if the right hand side of (\ref{eqq2}) has the form (\ref{2pol}), then the constants
$k_1,k_2$ can be arbitrary, and $b_1,b_2$ satisfy the
system
\begin{equation}\label{2lin}
b_1'=a (b_1-b_2) (1-k_1), \qquad  b_2'=a (b_2-b_1) (1-k_2).
\end{equation}

In the case, when $k_1+k_2=2,$ system (\ref{2lin}) possesses the
solution $a=1; b_i=u+t_i,$ where $k_1=1+ 1/(t_1-t_2),\, k_2=1+
1/(t_2-t_1).$ It corresponds to Case 4 from Section 3 with $c_1\ne
0,\, c_2^2\ne 4 c_1 c_3$.  For the Kupershmidt chain (see Example
2),
$$
\frac{R}{Q}= \frac{k_1}{x-b_1},
$$
$a=1$ and $b_1=u+t_1.$

For Case 2 with $c_1\ne 0,\, c_2^2\ne 4 c_1 c_3$ we have
(\ref{2pol}), (\ref{2lin}) and $k_1+k_2=1.$ Under the latter
condition, (\ref{2lin}) possesses the solution $a=1/u; \, b_i=t_i
u,$ where $k_1=t_2/(t_2-t_1), k_2=t_1/(t_1-t_2).$

Case 1 corresponds to a constant solution of system (\ref{2lin}),
which exists for $k_1=k_2=1.$ It was already mentioned in Section
3 that Case 3 is equivalent to Case 1.

\section{Conclusion}

We apply the method of hydrodynamic reductions to classify
integrable Vlasov-type equations of the form (\ref{maineq}), (\ref{psi}).
In this paper the simplest case of one function $U(x,t)$ is
completely analyzed. In the next paper we are going
to solve a more complicated problem of classification of integrable
Vlasov-type equations in the case of two functions
$U_1(x,t),U_2(x,t).$ It turns out that there exist several essentially
different classes of integrable functions $\psi(U_1,U_2,p)$.
One of such classes corresponds to two-component (2+1)-dimensional
hydrodynamic-type systems. This class was constructed in the paper \cite{odes}.
Note that examples of integrable
functions $\psi(U_1,\dots,U_n, p)$ appeared earlier in other
papers, where other approaches were used. In particular,
functions $\psi,$ associated with algebraic curves of an arbitrary
genus, were constructed in \cite{kr4}.
An integrable function $\psi(U_1,\dots,U_n, p)$ was constructed from any
$n$-dimensional Frobenius manifold in \cite{dub}.

As it was shown in \cite{Gibt}, equation (\ref{qq0}) in the case  of the
Benney chain (see Example 1) is nothing but the Loewner equation well
known in the theory of conformal mappings. The
results obtained in our paper can be of interest in connection with the so-called Laplacian growth problem
(see \cite{lg}
and references therein). Moreover, each integrable case
leads to an integrable hydrodynamic chain similar to the Benney chain (see
Example 1.) If the range of the discrete variable $k$ is the set of all integers,  the
corresponding hydrodynamic chains can be constructed rather easily. However, the
problem of a "right"  truncation of such chains to the set of non-negative values of $k$ is not trivial.
We are going to write a separate paper on the subject.

\vskip.3cm \noindent {\bf Acknowledgments.} Authors thank B.A.
Dubrovin and E.V. Ferapontov for fruitful discussions. We are
grateful to the ESF Research Network MISGAM for partial financial
support of the ISLAND-3 conference, where the work on this paper was
initiated. V.S. thanks IHES and A.O. thanks MPIM for hospitality and
financial support. M.P. was partially supported by the
Russian-Italian Research Project 06-01-92053. V.S. was partially
supported by the RFBR grants 08-01-461 and NS 1716.2003.1.

\end{document}